\def\qa{~\times 10^{44}}
\def\qb{~\times 10^{50}}
\def\try{t_{R}}
\def\et{~\times~10^{15}~sec}
\def\as1{\overline{A_{1}}}
\def\ea{~\times~10^{19}}
\def\eb{~\times~10^{14}}
\def\bs1{\overline{B_{1}}}
\def\oma{\omega_{\alpha}}
\def\tg{\tilde t}
\def\tgh{\tilde t^{2/3}}
\def\tgm{\tilde t^{8/3}}
\def\ter{\tilde {T_{e}}}
\def\terd{\tilde {T_{e}}^{-1/2}}
\def\tery{\tilde {T_{e}}^{1/2}}
\def\tgd{\tilde t^{2}}

\def\nib{n_{PBH}}
\def\omnor{\omega_{NOR}}

\def\nep{n_{e}}
\def\ned{n_{e}^{2}}
\def\zre{z_{R}}
\def\hab{\hbar}
\def\tred{10^{13}}
\def\cosi{\hbar c^{3}}
\def\cosq{\hbar  c^{4}}
\def\tri{t_{r}^{-1}}
\def\ppi{t_{pi}^{-1}}
\def\tco{t_{coll}^{-1}}
\def\tee{T_{e}}

\def\part{\partial t}

\def\paro{\partial \omega}

\def\duto{{2\over 3\tilde t}}

\def\suto{{8\over 3\tilde t}}
\def\rax{\dot R}
\def\scal{{\rax\over R}}
\def\num{n_{\gamma}}
\def\nure{n_{\gamma R}}
\def\nupi{n_{\gamma PI}}
\def\nuso{n_{\gamma PR}}
\def\nugg{n_{\gamma PR}}
\def\parn{\partial n_{\gamma}(\omega,t)}
\def\parnum{\partial n_{\gamma}}
\def\parp{\partial\nuso}
\def\park{\partial\nugg}
\centerline{\bf THE INFLUENCE OF QUARKS AND GLUONS JETS}
\centerline{\bf COMING FROM PRIMORDIAL BLACK HOLES}
\centerline{\bf ON THE REIONIZATION OF THE UNIVERSE.}
\vskip 1cm
\centerline{\bf{Marina Gibilisco}}\vskip 1mm
\centerline{\it Queen Mary and Westfield College,}\vskip 1mm
\centerline{\it Astronomy Unit, School of Mathematical Sciences,}\vskip 1mm
\centerline{\it Mile End Road, London E1, 4NS,}\vskip 1mm
\centerline{\it and Universit\'a degli Studi di Milano,}\vskip 1mm
\centerline{\it Via Celoria 16, 20133 Milano, Italy.}
\vskip 6mm
\centerline{{\it PACS codes: 9760L, 9870V, 9880D, 9530J}}
\vskip 5mm
{\it Submitted to Annals of Physics, February 1996}
\vskip 5mm
\centerline{\bf Abstract:} 
\vskip 7mm
\noindent In a previous work, I discussed the effect of the 
primordial black holes (PBHs) quantum evaporation on the reionization of the
Universe at small redshifts ($z \leq 60$): in principle, the photons 
emitted during the evaporation of such objects could
drive a new ionization for the Universe after the
recombination epoch ($z\sim 1200$); this reionization 
process should happen during the last stages of the PBHs life, 
when they totally evaporate and emit a lot of massive and non 
massive particles.
The critical mass of a black hole whose 
lifetime is equal to the present age of the Universe is 
$\sim 4.4\times 10^{14}~h^{-0.3}~g$: thus, PBHs having a mass 
$M~\sim~10^{14}~g$ are the ideal candidates to induce
a reionization at small redshifts.

While in my previous study I considered an exact blackbody 
photon emission spectrum, here I will adopt
a more realistic one, taking into account the quark and gluons jets emission
through the contribution of a known fragmentation function.
When the BH temperature rises above the QCD confinement 
scale $\Lambda_{QCD}$, one should expect an important contribution from
quarks and gluons emission in the form of jets. 
In this paper I also improved my analysis by considering 
without any approximation
the cooling effects in the plasma temperature evolution;
as a result, I obtained a satisfactory 
``late and sudden'' reionization process, characterised by 
a very well controlled rise of the 
plasma temperature: the plasma heating is not so high
to induce a strong distortion of the
CBR spectrum, in agreement with the recent FIRAS upper limit on the 
comptonization parameter, $y_{c}<2.5 \times 10^{-5}$.
\vfill\eject
\centerline{\bf 1. INTRODUCTION}
\vskip 7mm
The possibility that the Universe has been reionized after the recombination
is strongly suggested by many experimental evidences as, for instance,
the Gunn Peterson test [1]: the
absence of the Lyman-$\alpha$ absorption line in the spectrum 
of high redshift quasars and the unexpected, low density of neutral hydrogen
($n_{H I}< 10^{-11}~cm^{-3}$)
in the intergalactic medium (IGM) suggest this hypothesis.

The causes of such a reionization for the Universe are unclear: the
collisional ionization of the IGM by cosmic rays [2] or by 
far-UV photons produced by quasars [3] have been proposed as 
a possible source but, in general,
a complete theory describing this phenomenon is not yet available.

An useful classification of the theoretical models presently 
known can be found in Ref. [4], where one distinguishes 
''late and sudden'' (LS) models (where typically the reionization
happens at $\zre < 60$ and it is very fast)
and ''early and gradual'' (EG) models, including, for instance,
those mechanisms that assume as a ionizing source 
the UV radiation from decaying, massive neutrinos [5].

In Ref. [6] I discussed a possible LS reionization mechanism
based on the quantum evaporation of primordial black holes:
as Hawking proved [7], the evolution of
a black hole involves a continuous loss of mass, 
a phenomenon ending with the complete evaporation of such an object. 
This is a typical quantum 
effect: in fact, in a classical sense, a black hole 
can only absorb particles but the quantum mechanical handling of the 
matter fields in a curved space-time involves some ambiguities in the 
decomposition of 
the field operators into positive and negative frequency components.
Thus, the annihilation and 
creation operators finally work on a vacuum state that is different 
in regions of the space-time having a different 
curvature [7] and, as a consequence, 
the creation of particles at the expense of the black hole mass 
becomes possible.

As I proved in Ref. [6],
the photons emitted during the evaporation process may induce the
reionization of the Universe; however, in order that such a
mechanism works, one should be careful the emitted photons
do not heat in an excessive way 
the background plasma. Such a heating should contradict the 
FIRAS upper limit on the comptonization parameter,
$y_{c}<2.5 \times 10^{-5}$ [8], a value
that definitely rules out the possibility of 
a relevant distortion of the CBR blackbody spectrum via the Sunyaev
Zel'dovich effect [9]. 
The proportionality of $y_{c}$ to the plasma temperature
clearly excludes the possibility that some processes in the past might 
have strongly heated the background plasma: if this is the case,
very effective cooling mechanisms should have operated to cut off the
temperature.

Due to some numerical problems, in Ref. [6] I solved in an approximate 
way the coupled differential 
equations system giving the evolution of the ionization degree $x$ and of the 
plasma temperature $T_{e}$: thus
the collisional and excitation cooling terms were
underestimated.

In the following sections,
I will present an improved calculation that solves the equation system 
taking into account in an exact way all the cooling terms; moreover,
I will consider a modified, non blackbody photon emission spectrum
which contains also the effect of quantum jets produced by PBHs at high 
temperatures ($T > \Lambda_{QCD}$): the importance of such a jets emission
has been put in evidence in Ref. [10] and may have some influence on the
reionization process.
 
As a consequence of these improvements, I obtained a satisfactory evolution
of both the ionization degree $x$ and the plasma temperature $T_{e}$:
in particular, the plasma heating is less relevant than in my previous 
study. The modification of the emission spectrum has the 
consequence that a lower density of primordial black holes is necessary
to produce a significant reionization, because in this case the photons 
coming from the secondary decays of hadrons add to the ones produced by
the direct evaporation;
a rough estimate of the present density parameter $\Omega_{PBH}$
inferred from this calculation is also given.

Future improvements are still possible: for instance, one may 
investigate how the production of massive
particles, in particular electrons, might modify the background 
plasma and possibly influence the Inverse Compton scattering processes.
Note however that the emission of charged 
particles represents a problem not immediately solvable because, in this case,
one should also study the charge evolution of the black hole:
from the theory, we know that the loss of charge and angular momentum 
of a charged, rotating black hole is very fast [11] but, as 
pointed out in Ref. [12], large, electrically charged 
BHs have a very interesting evolution towards the 
extreme Reissner-Nordstr\"om limit.
During this evolution, very peculiar thermodynamic
properties appear and 
various phase transitions may income during the evaporation process.
The consequences of this 
particular evolution are a lifetime many order
of magnitude larger than the one of a common Schwarzschild black hole
and an emission spectrum explicitly containing the charge evolution [13]:
probably, in these conditions, 
my analysis of the reionization 
should be radically modified and the conclusions I drew for the
uncharged case cannot be simply extended without a deep examination of 
the processes involved in the BHs discharge.

The consideration of the 
jets emission represents the first step toward a deeper 
understanding of the problem: in this paper, I will compare the 
characteristics of the reionization process both in the case of an exact
Hawking spectrum and in the one 
when a jet fragmentation function is introduced in order to 
consider also the secondary photon emission from the hadron decays.

\vskip 7mm
This paper is structured as follows: in Sec. 2, I will discuss the general
properties of primordial black holes, in particular the characteristic of their
quantum evaporation with and without jets emission;
in Sec. 3, I will recall the main formulas and results of my previous work,
in particular discussing the differential equations system for the variables
$x$ and $T_{e}$; in Sec. 4 I will present an improved 
numerical solution method that uses the new photon spectrum and does not
involve any approximations for the cooling terms; in such a way, 
a noteworthy improvement in the plasma temperature behaviour is obtained.

An estimate of the density of PBHs one needs in order to have a significant
reionization for the Universe is also given: it corresponds to a 
present density parameter $\Omega_{PBH}$ ranging from 
$1.12\times 10^{-12}$ to $1.65\times 10^{-8}$, depending on the PBHs formation
time one assumes.

Finally,
in sec. 5 I will discuss the new results in comparison with the ones
obtained in Ref. [6]; the conclusion is 
that the proposed model of a reionization induced by 
quantum evaporation of PBHs seems to work 
without producing an excessive plasma heating and seems to be 
effective in the range
of reionization redshifts [15, 60] here considered. Further extensions
to higher $\zre$ values are also possible.
\vskip 7mm
\centerline{{\bf 2. THE PHYSICS OF THE PRIMORDIAL BLACK HOLES:}}
\centerline{{\bf QUANTUM EVAPORATION AND PHOTON}} 
\centerline{{\bf EMISSION SPECTRUM.}}
\vskip 7mm
The fundamental work of Hawking about black holes physics [7] puts
in evidence many interesting properties for such objects,
in particular their peculiar connections with the general 
thermodynamics.

The creation of a black hole becomes possible when a mass contracts 
to a size less 
than its gravitational radius: typically, this is the case for large mass stars
at the end of their evolution; however,
for objects having smaller masses, a compression to huge 
densities is necessary in order to create a black hole; in the
same time, the large pressure forces counteracting the compression
should also be overcome.

For this reason, the formation of black holes having a small mass 
($M << M_{\odot}$) is very improbable in the contemporary Universe:
however, Hawking's quantum evaporation phenomena 
are just important for small-mass black holes, the 
blackbody temperature of the emission being:
$$
kT~=~{{\hab c^{3}}\over {8\pi GM}}~\sim~
1.06~\Big[~{M\over {\tred ~g}}~\Big]^{-1}
~GeV.\eqno(2.1)
$$
Therefore, our interest is mainly fixed upon the so-called {\it primordial}
black holes (PBHs): at the early stages of the cosmological expansion,
the density of the matter was really huge, thus 
enabling the formation of small-mass BHs.

As stressed in Ref.
[14], small perturbations in a homogeneous, isotropic, hot Universe 
would not be able to produce relevant inhomogeneities without the
contemporary presence of large fluctuations in the gravitational field;
the possibility of the creation of a black hole just appears when the
quantity $l=ct$ ($t$ is the time elapsed since the Big Bang)
grows to a value in the order of the metric perturbation size and its
resulting
mass will be equal to the mass contained in a volume $l^{3}$ at the time $t$
[14]. The r\^ole of primordial density perturbations in the PBHs 
formation processes is discussed in a more extensive way in [15]; in
particular, a study of the effect of 
the adiabatic inflaton quantum fluctuations 
in chaotic models can be found in [16]; other possible formation mechanisms 
might be a cosmological phase transition involving bubbles collision [17], 
a softening of the Universe equation of state [18] or a collapse of 
cosmic strings [19].

In this work, my interest is mainly 
concentrated on the evaporation of PBHs
that form during the first stages of the expansion of the Universe:
a choice of the possible birth times will be done in the next sections
in connection with the calculation of the PBHs initial density we need 
in order to have an appreciable reionization. 

The kind of the emitted particles
obviously depends on the blackbody temperature of the PBHs: following 
eq. (2.1), the mass loss during the quantum evaporation of a BH 
makes it hotter and hotter, thus enabling it to produce 
more and more massive particles. 
Typically, BHs having a mass larger than $10^{17}~g$
emit massless particles only, like photons, neutrinos
and, may be, gravitons; electron production should be expected 
by BHs whose mass 
is in a range $10^{15}~g < M <10^{17}~g$ and muon production for a range 
$10^{14}~g < M <10^{15}~g$. 

Here I will consider PBHs that survive 
till the post-recombination epoch, thus having 
a mass $M\sim O(10^{14})~g$, 
near to the critical mass,
$M_{c}\sim 4.4\times 10^{14}~h^{-0.3}~g$.
These PBHs should reionize
the Universe at a redshift $\zre$ that corresponds to the time 
of their complete evaporation.

The initial mass $M_{i}$ is connected to the lifetime of a PBH by 
the following formula [20]:                                        
$$
t_{evap}~\sim~1.19\times 10^{3}~{G^{2}M_{i}^{3}\over \cosq ~f(M_{i})}~\sim~
6.24\times 10^{-27}~f(M_{i})^{-1}~M_{i}^{3}~sec,\eqno(2.2)
$$
where the function $f(M)$ contains 
the contributions of the different species
of particles and it is normalized to the unit for very massive
($M \geq 10^{17}~g$) BHs, emitting massless particles only; a suitable
way to express these contributions is given by the following formula [20]:
$$
f(M)~=~1.569~+~0.569~\Bigg[~exp\Big[~{-M\over {4.53\cdot 10^{14}}}~\Big]_{\mu}~
+~6~exp\Big[~{-M\over {1.60\cdot 10^{14}}}~\Big]_{u,d}~+~
$$
$$
+~3~exp\Big[~{-M\over {9.60\cdot 10^{13}}}~\Big]_{s}
~+~3~exp\Big[~{-M\over {2.56\cdot 10^{13}}}~\Big]_{c}~
+~exp\Big[~{-M\over {2.68\cdot 10^{13}}}~\Big]_{\tau}~+
$$
$$
~+~3~exp\Big[~{-M\over {9.07\cdot 10^{12}}}~\Big]_{b}
~+~3~exp \Big [~{-M\over {0.48\cdot 10^{12}}}~\Big ]_{t}~ \Bigg]~+~
$$
$$
+~0.963\Big[~exp\Big[~{-M\over {1.10\cdot 10^{14}}}~\Big]_{gluons} 
\Big ].\eqno(2.3)
$$
In eq. (2.3) the first addendum in the right-hand side expresses the 
contribution of 
electrons, positrons, photons and neutrinos;
heavier particles contributions
are considered in the remaining terms, following their relative importance;
the factor 3 takes into account the presence of the color charge
for quarks and the denominators in the exponential terms are defined as 
the product $\beta_{sj}M_{j}$, where $M_{j}$ is the mass of a black hole 
whose temperature is equal to the rest mass $\mu_{j}$ of the $j$ species and 
$\beta_{sj}$ is a spin-dependent factor defined [20] in such a way the energy 
of a BH having $M=\beta_{sj}M_{j}$ has a peak at $\mu_{j}$.

Eq. (2.3) is fundamental in the determination of the mass evolution of a PBH 
that evaporates: in fact, in Refs. [10], [20], Carr, Mac Gibbon
and Webber show that the quarks and gluons emission processes 
should not be neglected at the stages when the BH mass falls 
below $10^{14}~g$ and the temperature $T$ becomes larger than the confinement 
scale $\Lambda_{QCD}$: at this time one should expect a copious production of 
jets and this is the case we are interested in, due 
to the particular range of initial masses
chosen for the PBHs.

Through the function $f(M)$, eq. (2.3),
we can take into account the quarks/gluons emission processes:
the mass evolution is given by:
$$
{dM\over dt}~=~-\sum_{j} {1\over 2\pi\hab}~\int \Gamma_{j}~\Bigg[~exp
~\Big[~{8\pi GQM\over \cosi}~\Big]~-~(-1)^{2s_{j}}~\Bigg]^{-1}\times
{Q~dQ\over c^{2}};\eqno(2.4)
$$
eq. (2.4) means that the emission of a 
parent particle $j$ with total energy $Q$ decreases the BH mass by $Q/c^{2}$;
here $\Gamma_{j}$ is the absorption probability for the $j$ particle 
having a spin $s_{j}$ [21] and 
a sum on all the emitted species has been performed [20].
After the integration over the energy $Q$, eq. (2.4) 
can be rewritten as [20]:
$$
{dM\over dt}~=~-5.34\times 10^{25}~f(M)~M^{-2}~g~sec^{-1}\eqno(2.5)
$$
Now, the Hawking emission rate of particles having an energy in the range 
$(E, E+dE)$ from a black hole having an angular velocity $\omega$,
an electric potential $\phi$ and a surface gravity $\kappa$ is [7]:
$$
{dN\over dt}~=~{\Gamma ~dE\over 2\pi\hbar }~\Bigg[~exp~\Bigg(~
{{E-n\hbar  \omega-e\phi}\over{\hbar \kappa/2\pi c}}~\Bigg)~\pm~1~\Bigg]^{-1},
\eqno(2.6)
$$
where the signs $\pm$ respectively refer to fermions and bosons
and $\Gamma$ is the absorption probability of the emitted species: for photons,
it reads as [21]:
$$
\Gamma_{s=1}~=~{4A\over 9\pi}~\Big({M\over M_{PL}}\Big)^{2}~
\Big({\omega\over \omega_{PL}}\Big)^{4};\eqno(2.7)
$$
here $A$ is the surface area of the BH
and the Planck mass and energy have been introduced in order to work
with dimensionless quantities as in Ref. [21].

In most cases, one assumes that 
the charge and the angular momentum of a black hole are negligible
because, as Page proved [13], their loss through quantum evaporation
happens on a time scale shorter than the one characterizing the mass loss;
this assumption is certainly reasonable but one should also note that
in some way it may be relevant to determine the influence 
of the black hole charge evolution (in particular studying 
the discharge process) in connection with
the modification of the electron background; thus, one should rather
follow the approach suggested in ref. [12].

In Ref. [6] I studied the reionization of the Universe induced 
by quantum evaporation of PBHs by using a photon emission spectrum as the one 
of eq. (2.6), i.e. an exact blackbody spectrum;
this assumption enabled me just to 
test if the proposed model of reionization might work at least under 
very general conditions. After this satisfactory test, I will 
now improve my previous 
calculation by studying the effect of the presence of a jet 
fragmentation function contribution in the emission spectrum:
then, eq. (2.6) should be rewritten as [22]:
$$
{dN_{x}\over dtdE}~=~\sum_{j}~\int^{+\infty}_{0}~{
\Gamma_{j}(Q,T) \over 2\pi\hbar }~\Big(~exp{Q\over T}\pm1~\Big)^{-1}~
{dg_{jx}(Q,E)\over dE}~dQ;\eqno(2.8)
$$
here $x$ and $j$ respectively refer to the final and the directly emitted 
particles and the last factor, containing the 
fragmentation function $g_{jx}$, expresses
the number of particles with energy in the range $(E, E+dE)$ coming
from a jet having an energy equal to $Q$: namely, it reads as [22]:
$$
{dg_{jx}(Q,E)\over dE}~=~{1\over E}~\Bigg(~1-{E\over Q}~\Bigg)^{2m-1}~
\theta(E-km_{h}c^{2}),\eqno(2.9)
$$
where $m_{h}$ is the hadron mass, $k$ is a constant $O(1)$
and $m$ is an index equal to 1 for mesons and 2 for baryons. 

After determining the 
dominant contribution to the integral over $Q$ and 
summing over the final states, one can 
approximate eq. (2.8) by the following formulas [22]:
$$
{dN\over dtdE}~\sim~E^{2}~exp \Big({-E\over T}\Big) ~~~~~~~~~~~~~for
~~E >> T~~~~~~ ~Q\sim E, \eqno(2.10a)           
$$
$$
{dN\over dtdE}~~\sim~ E^{-1}~~~~~~~~~~~~~~~~~~~~ for ~~T\sim E >> m_{h}~~
~~~~Q\sim T,\eqno(2.10b)
$$
$$
{dN\over dtdE}~\sim~{dg \over dE}~~~~~~~~~~~~~~~~~~~~~~~~~~for 
~~E\sim m_{h} << T~~~~~~
Q\sim m_{h},\eqno(2.10c)
$$
where the different expressions refer to the specified $Q$-value that
dominates. A more exhaustive discussion of the characteristics of the spectrum 
given by eqs. (2.10a), (2.10b) and (2.10c) can be found in Ref. [22].

In the next Section, I will examine the problem of the reionization
of the Universe by considering as a source of the ionizing photons the 
quantum evaporation of PBHs in presence of quarks and gluons jets.
\vskip 7mm
\centerline{{\bf 3. THE TIME EVOLUTION OF THE IONIZATION DEGREE}}
\centerline{{\bf AND PLASMA TEMPERATURE FOR A REIONIZED UNIVERSE.}}
\vskip 7mm
As I discussed in Ref. [6], the basic equations that control the 
ionization degree $x$ and the plasma temperature $T_{e}$ evolution
with the time are [23]:
$$
{dx\over dt}~=~\ppi~+~\tco~-~\tri,\eqno(3.1)
$$
and 
$$
{d\tee\over dt}~=~-2~\scal~\tee~-~{\tee\over (1+x)}~{dx\over dt}
~+~{2\over 3(1+x)}~(\Gamma -\Lambda),\eqno(3.2)
$$
where $\ppi ,~\tco ,~\tri$ are, respectively, the photoionization, 
the collisional and the recombination rates, given by [23]:
$$
t_{pi}^{-1}~=~{8\pi m_{e}\alpha^{5}\over 3\sqrt{3}}~(1-x)~{1\over \pi^{2}}~
\int^{\omega max}_{\omega min}~d\omega ~{\num\over \omega^{3}};\eqno(3.3)
$$
$$
t_{r}^{-1}~\sim~10^{-13}~sec^{-1}~\Big ({\Delta\over T_{e}}\Big)^{1/ 2}
~x^{2}~\Big ({\Omega_{b}~h^{2}\over 0.025}\Big)~\Big({{1+z}\over 200}\Big)
^{3}~~~~~~~~~~~~~~~~~~(\Delta= 1~Ry);\eqno(3.4)
$$
$$
t_{coll}^{-1}~=~6\times 10^{-8}~sec^{-1}~\Big ({T_{e}\over\Delta }\Big)^{1/ 2}
~e^{-\Delta / T_{e}}~x~(1-x)
~\Big ({\Omega_{b}~h^{2}\over 0.025}\Big)~\Big({{1+z}\over 200}\Big)
^{3}.\eqno(3.5)
$$
Eq. (3.1) expresses the combined effects of the
photoionization and the recombination and the influence of the collisional 
interactions putting the neutral
hydrogen in an excited or ionized state.
In eq. (3.2), the heating $\Gamma$ mainly comes from the 
photoionization process and reads as
$$
\Gamma~=~\Gamma_{pi}~=~{8\pi m_{e}\alpha^{5}\over 3\sqrt{3}}~(1-x)~
{1\over \pi^{2}}~
\int^{\omega max}_{\omega min}~d\omega ~{(\omega-\Delta)\over \omega^{3}}~
\num;\eqno(3.6)
$$
on the contrary, 
the cooling $\Lambda$ takes into account a lot of contributions
having a relevant importance in order to limit the rise of the plasma 
temperature induced by 
the quantum evaporation of PBHs. These contributions 
are due to the recombination, collisional and excitation processes,
Compton scattering and to the adiabatic 
cooling due to the expansion of the Universe.
The total cooling reads as
$$
\Lambda~=~\Lambda_{r}~+~
\Lambda_{exc}~+~\Lambda_{coll}~+~\Lambda_{Compt}~+~
\Lambda_{exp},
\eqno(3.7)
$$
where, respectively:
$$
\Lambda_{r}~=~2\times 10^{-21}~GeV~sec^{-1}~\Big ({T_{e}\over\Delta }\Big)
^{1/ 2}~x^{2}~
~\Big ({\Omega_{b}~h^{2}\over 0.025}\Big)~\Big({{1+z}\over 200}\Big)
^{3};\eqno(3.8)
$$
$$
\Lambda_{exc}~=~10^{-15}~GeV~sec^{-1}~e^{-3\Delta / 4T_{e}}~
~\Big ({T_{e}\over\Delta }\Big)^{1/ 2}~x~(1-x)
~\Big ({\Omega_{b}~h^{2}\over 0.025}\Big)~\Big({{1+z}\over 200}\Big)
^{3};\eqno(3.9)
$$
$$
\Lambda_{coll}~=~8\times 10^{-16}~GeV~sec^{-1}~\Big ({T_{e}\over\Delta }\Big)
^{1/ 2}~                                   
e^{-\Delta / T_{e}}~x~(1-x)
~\Big ({\Omega_{b}~h^{2}\over 0.025}\Big)~\Big({{1+z}\over 200}\Big)
^{3};\eqno(3.10)
$$
$$
\Lambda_{Compt}~=~10^{-19}~Gev~sec^{-1}~x~\Big({{T_{e}- T_{CMB}}\over\Delta}
\Big)~\Big({{1+z}\over 200}\Big)^{4};\eqno(3.11)
$$
$$
\Lambda_{exp}~=~2.\times 10^{-22}~GeV~sec^{-1}~\Big({T_{e}\over \Delta}\Big)~
(1+x)~\Big({{1+z}\over 200}\Big)^{1.5}.\eqno(3.12)
$$
($T_{CMB}$ is the Cosmic Microwave Background temperature).
Both the equations (3.3) and (3.6) explicitly contain the photon 
number density $n_{\gamma}$: the time evolution of this 
variable is quite difficult to establish because it depends on 
many processes not simply correlated;
the best way to describe this evolution is through a differential equation 
that reads as follows [6]:
$$
{\parn\over\part}~+~\scal~{\parn\over\paro}~\omnor~-~2~\scal~
{\num (\omega, t)\over\omega}~\omnor~=
$$
$$
=~\Big(~{d\nure\over dt}
-{d\nupi\over dt}~\Big)~+~
{2\over \omega}~{d\omega\over dt}~[{\nupi - \nure}]~+~{\park\over
{\paro\part}}~\omnor;\eqno(3.13)
$$
here $\nupi,~\nure$ are the photon number densities respectively involved
in the processes of photoionization and recombination and $\omega_{NOR}$
is a normalization factor, equal to $10^{-6}~GeV$: this normalization
assures one is working with dimensionless
quantities in the numerical routine that solves the differential equation
system.
The last term in eq. (3.13)
is the contribution of the photon source, in our case
a number $n_{PBH}$ of primordial black holes that evaporate.

From the basic work of Peebles [24], the first term on the right-hand side
of eq. (3.14) has the following form:
$$
\Big(~{d\nure\over dt}-{d\nupi\over dt}~\Big)~=~-{d\nep \over dt}~=~
C~\Big(~\alpha_{e}\ned~-~\beta_{e}n_{1s}e^{-\oma /T}~\Big)~-~{\nep\over n}
~{dn\over dt}.\eqno(3.14)
$$
Explicitly,
$\alpha_{e}\ned$ is the rate of recombination to excited states,
ignoring the recombination direct to the ground state,
$\alpha_{e}=\langle \sigma\nu\rangle $ is the recombination coefficient 
and $\beta_{e}n_{1s}e^{-\oma /T}$ expresses the rate of 
photoionization; $\beta_{e}$ is a constant 
proportional to $\alpha_{e}$, $n_{1s}$ is the population
of the $1s$ state of the hydrogen atom, $\omega_{\alpha}$ is the 
energy corresponding to the transition $L_{\alpha}$, $T$ the radiation
temperature and $n$ is the nucleon number density. Finally, $C$ is a 
reduction factor which takes into account the effect of the $L \alpha$
resonance photons.
All the numerical values of these quantities can be found in Ref. [25];
really, we do not mind this contribution because it proves 
to be negligible with respect to the difference $\nupi - \nure$.

Now, as in Ref. [6], I put 
$$
\num~=~\nuso~-\nupi~+~\nure , \eqno(3.15)
$$
and with the same substitution performed in [6] I obtain:
$$
{\parn\over\part}~+~2
~{\num\over\omega}~\Big[-~\scal~\omnor +{d\omega\over dt}~\Big]~+~\scal~
{\parn\over\paro}~\omnor~=~
$$
$$
=~C~\Big 
(~\alpha_{e}x^{2}n^{2}~-~\beta_{e}n_{1s}e^{-\oma /T}
~\Big)~-~x~{dn\over dt}
~+~{\park\over
{\paro\part}}~\Big(~{2\over \omega}{d\omega\over dt}~+~\omnor~\Big).
\eqno(3.16)
$$
At this point my present analysis 
of the problem differs from the approach I followed 
in Ref. [6]: in fact, here I will take as a source term in eq. (3.13) 
the photon emission spectrum corrected by using a jet
fragmentation function (eqs. (2.8), (2.10a), (2.10b), (2.10c))
while previously I used an exact 
Hawking photon spectrum, eq. (2.6).
\vskip 7mm
\centerline{{\bf 4. THE NUMERICAL SOLUTION OF THE DIFFERENTIAL}}
\centerline{{\bf EQUATION SYSTEM.}}
\vskip 7mm
I will start the present analysis of the reionization mechanism by 
evaluating the total photon number density coming from the quantum 
evaporation of PBHs: eqs. (2.10a), (2.10b) and (2.10c) give
the emission rate for various 
values of the energy characterizing the emitted particles. 

As I stressed in Ref.
[6], I would take into account only the contribution of those photons that 
may really reionize the Universe, i.e., I want to select 
the components of the emission spectrum really effective 
by comparing the time scale characterizing the 
photon interactions with the expansion time for the Universe. 
The following condition should be satisfied:
$$
t_{int}<t_{exp}.\eqno(4.1)
$$
As a consequence of the condition (4.1), the latest time at which the 
photon interactions may be relevant is [10]:
$$
t_{free}~=(~n_{X}~\sigma~)^{-1}.\eqno(4.2)
$$
($n_{X}=$ number density of the background particles, $\sigma =$ interaction 
cross-section). 

Among the possible interaction processes, I
want to consider the ionization and the recombination: the range
of energy where these processes are dominating is typically
$E_{\gamma}\leq 14~KeV$ but, indeed, the maximum allowed energy for photons 
having a r\^ole in the reionization of the Universe
is further constricted by the condition (4.1);
in the pregalactic, 
matter-dominated era, a photon is affected by ionization losses only before a 
redshift [10]:
$$
1~+~z_{free}~\sim~1.0\times 10^{19}~h^{-2}~\Big({0.2\over\Omega_{p}}\Big)
~E^{3.5},\eqno(4.3)
$$
where $\Omega_{p}\sim 0.2$ is the present density parameter for protons, 
$h=0.5$ and the energy is expressed in $GeV$.

Eq. (4.3) can be inversely used in order to determine $E_{max}$
at a particular value of $z$, corresponding to the reionization redshift
$\zre$:
in tab. 1 I listed the values I obtained, typically smaller than
$8~KeV$ for a late reionization with $\zre < 60$.

Coming back to the eqs. (2.10a), (2.10b) and (2.10c), 
due to the 
upper limit we have for the photon energy, the form of the spectrum
is the one expressed by eq. (2.10c), holding in the case $E << T$.

Taking the
jet fragmentation function as in eq. (2.9) and separating the mesons and 
baryons contributions, I can write:
$$
{\parnum\over \paro\part}|_{Tot}~=~{\parnum\over \paro\part}|_{Mes}~+~
{\parnum\over \paro\part}|_{Bar},\eqno(4.4)
$$
(for convenience, in the following discussion the energy of the 
final photons will be called $\omega$).

For $Q=m_{hadr}\sim 300~MeV$, eq. (4.4) simply reads:
$$
{\parnum\over \paro\part}|_{tot}~=~{1\over \omega}~\Big(1-{\omega\over Q}
\Big)~+~
{1\over \omega}~\Big(1-{\omega\over Q}\Big)^{3},\eqno(4.5)
$$
the dominating contribution being the mesonic one. 

Turning now to the solution of the photon density equation, it is useful to
adopt dimensionless rescaled variables, 
$\tilde\omega=\omega / \omnor$ with $\omnor=10^{-6}~GeV$,
$\tilde M= M / 10^{14}~g$ and 
$\tilde t=t / 10^{15}~sec$ and to write eq. (3.16) in the form
$$
{\parnum\over \partial \tilde t}~+~{2\over 3\tilde t}~{\parnum\over
\partial\tilde\omega}~=~
{4\over 3\tilde t}~{n_{\gamma}\over \tilde \omega}~-~2
{n_{\gamma}\over \tilde \omega}~{d\tilde \omega\over d\tilde t}~+~
{\parp\over\partial\tilde t\partial\tilde\omega}~
\Bigg(~1+{2\times 10^{-9}\over \tilde\omega}~
{d\tilde \omega\over d\tilde t}~\Bigg).\eqno(4.6)
$$
In eq. (4.6) the constant $2\times 10^{-9}$ is a numerical factor coming from
the transformation to the new rescaled variables and all the negligible 
terms have been skipped away.

From the explicit calculation of eq. (4.5), one has:
$$
{\parp\over\partial\tilde t\partial\tilde\omega}~
~\sim~~{2\times 10^{15}\over\tilde \omega}.\eqno(4.7)
$$
Now, the general solution of eq. (4.6) can be found in an analytical way 
by using the method discussed in Ref. [6]: here, I would like just 
to recall that the general solution of a first order, linear partial
differential equation that reads as
$$                              
Pp~+~Qq~=~R,\eqno(4.8)
$$
($p={\partial z\over \partial x}$, $q={\partial z\over \partial y}$
and $P,~Q,~R$ are any function of the general variables $x,~y,~z$)
is  a function $F(u,v)$ that also solves [26] the equations
$$
|{dx\over P}|~=~|{dy\over Q}|~=~|{dz\over R}|;\eqno(4.9)
$$
here $u(x,y,z)=c_{1}$, $v(x,y,z)=c_{2}$ are implicit functions and 
the constants $c_{1},~c_{2}$ should be fixed by imposing suitable boundary
conditions to the particular problem one is studying.

After the identification of 
the first variable, $x$, with the time $\tilde t$, the second one, $y$, with
the energy $\tilde \omega$ and finally $z$ with the photon number
density $n_{\gamma}$, I obtain:
$$
P~=~1~~~~~~~~~~~~~~~~~~~~~~~~~~Q~=~\duto,\eqno(4.10a)
$$
$$
R~=~\suto~{n_{\gamma}\over \tilde\omega}~+~
{2\times 10^{15}\over\tilde \omega};\eqno(4.10b)
$$
the main difference with the expression found in ref. [6] consists
in the modified source photon term. Contrarily to my previous 
analysis, here I do not examine the very particular case when one has
equilibrium between photoionization and recombination.

The left-hand side of the equality (4.9) gives:
$$
{d \tilde\omega \over d\tilde t}~=~-{2\over 3\tilde t},\eqno(4.11)
$$
as in Ref. [6], while for the photon density $n_{\gamma}$ I calculate
$$
{d n_{\gamma}\over d\tilde \omega}~=~{3 \tilde t\over 2}~R~=~
{1\over \tilde\omega}~\Big[~4 n_{\gamma}~+~
{3 \tilde t\over 2}~a~\Big],\eqno(4.12)
$$
where $a=2\times 10^{15}$. After solving the 
linear equation (4.12) one finds:
$$
n_{\gamma}(\tilde\omega)~=~-{3\over 8}a~\tilde t~+~c_{2}~\tilde\omega^{4};
\eqno(4.13)
$$
the integration constant $c_{2}$ is fixed by imposing that the final 
number of emitted photon is zero for $t=t_{0}$:
$$
c~=~{3\over 8}~a~{\tilde t_{0}\over \tilde \omega_{0}^{4}},
\eqno(4.14)
$$
where $\tilde t_{0}=t_{0}/10^{15}~sec$ and a good choice for
$\tilde\omega_{0}$ should be done in such a way one has a 
positive photon density at all times; from such a constraint, 
one obtains $\tilde\omega_{0}~\sim~\tilde\omega_{max}/12$;
moreover, a lower limit on the photons energy comes from 
the request that the emission process is not less than
the recombination losses: this minimum
value is equal to $0.20~KeV$
while, as I discussed in the previous sections, the maximum energy
depends on the reionization redshift $z_{R}$.

I adopted a numerical Merson routine to 
solve eqs. (3.1), (3.2): note however that, although  Merson is a very 
powerful method, it fails if the photon number density 
evolution as expressed by eq. (4.14) 
is directly inserted in in eqs. (3.3), (3.6) for $t_{PI}^{-1}$ and
$\Gamma_{PI}$ . 
Really, Merson cannot 
control the contemporary evolution of two variables like the time and the 
photon density in the same equation; a solution is only possible 
by making some simplifying assumptions, namely:

\item{a)} the collision contribution $\tco$ in eq. (3.1) is neglected with 
respect to the photoionization one: this assumption is certainly 
acceptable if we suppose that the quantum evaporation phenomena 
is the most effective contribution to the photon number
density. In this way, Merson should not control terms having coefficients 
different by many order of magnitude.

\item{b)} In eqs. (3.3), (3.6) I took $\omega_{min}=0.20~KeV$, i.e.
the minimum of the energy previously obtained;
then I put:
$$
A_{1}(\tilde t)~=~\int^{\tilde\omega_{max}}_{\tilde\omega_{min}}~
d\tilde\omega~{n_{\gamma}(\tilde\omega)\over \tilde\omega^{3}};\eqno(4.15a)
$$            
$$
B_{1}(\tilde t)~=\omega_{NOR}~\int^{\tilde\omega_{max}}_{\tilde\omega_{min}}~
d\tilde\omega~(\tilde\omega~-~\Delta)~
{n_{\gamma}(\tilde\omega)\over \tilde\omega^{3}}.\eqno(4.15b)
$$            
Now, I insert in eqs. (3.1) and (3.2) the time-averaged values 
$\overline {A_{1}}, ~\overline {B_{1}}$ that I quoted in tab. 2 for various
reionization redshifts:
in such a way, the Merson routine 
must solve a constant coefficients equation system and no 
approximations are necessary in the
handling of the cooling terms of eq. (3.2).

On the contrary, in ref. [6] I tried to maintain the time dependence
in the functions $A_{1},~B_{1}$ but I was obliged to 
insert some cooling terms in an approximate form;
the present choice seems most suitable also because 
the use of an average number of photons in eq. (3.1), (3.2) probably 
does not represent an extreme approximation, due to our 
poor experimental knowledge
of the primeval PBH density parameter $\Omega_{PBH}$; 
further improvements in the mathematical solution of the system are 
under study.

As a result of this solution method, I am able to control
the rise in the plasma temperature $T_{e}$ while previously the model involved
a too large effect of plasma heating. 
The effectiveness of the cooling mechanism can disclose if 
the suppression due to the negative exponential terms in eqs. (3.9), (3.10), 
i.e. to the excitation / collisional terms, is avoided by choosing 
to start with a high plasma temperature $T_{e}$ corresponding 
to an instantaneous heating.

Remembering this assumptions, the couple of differential equations 
is written as follows:
$$
{dx\over d\tilde t}~=~C_{1}(\zre)~n_{PBH}~(1-x)~-~77.45~{x^{2}\over \tgd}~
\terd;\eqno(4.16a)                                        
$$
$$
{d\ter\over d\tg}~=~-1.33~{\ter\over\tg}~-~{\ter\over (1+x)}~{dx\over d\tg}
~+~{0.67\over (1+x)}\cdot \Bigg[~C_{2}(\zre)~n_{PBH}~(1-x)~-
$$
$$
-~113.40~
{x^{2}\over \tgd}~\tery~-~
5.69\times 10^{7}~x~(1-x)~{1\over \tgd}~\Big(1-{1020\over \ter}\Big)~\tery~-~
$$
$$
-~4.54\times 10^{7}~~x~(1-x)~{1\over \tgd}~\Big(1-{1360\over \ter}\Big)~\tery~-~
44.10~x~\Big(\ter - {1.27\over \tgh}\Big)~{1\over\tgm}~-~
$$
$$
-~2.13~(1+x)~
{\ter\over \tg}~\Bigg];\eqno(4.16b)
$$
here $\ter=T_{e} / 10^{-11}~GeV$.

In eq. (4.16a) I neglected the collisional term $\tco$ (as I explained above);
thus, the addends in eq. (4.16a) respectively correspond to $\ppi$, $\tri$
and the ones in brackets in eq. (4.16b) to $\Gamma_{pi}$, $\Lambda_{r}$,
$\Lambda_{exc}$, $\Lambda_{coll}$, $\Lambda_{Compt}$ and
$\Lambda_{exp}$. Finally, the coefficients $C_{1}(\zre),~C_{2}(\zre)$
are given by:
$$
C_{1}(\zre)~=~7.88\times 10^{24}~\overline A_{1}(\zre),\eqno(4.17a)
$$
$$
C_{2}(\zre)~=~7.88\times 10^{35}~\overline B_{1}(\zre),\eqno(4.17b)
$$
and they are listed in tab. 3 for various $\zre$ from 15 to 60.

In eqs. (4.16a) and (4.16b), the parameter $\nib$ is a
weight factor that represents the effective number of PBHs in 
the early Universe we need in order to produce an appreciable reionization 
effect; a suitable value for this free parameter is
$\nib =10^{-44}$;
an estimate in terms of an effective initial number density of PBHs 
corresponding to this value of $\nib$
can be roughly obtained by considering the following relations:
$$
\rho_{i}~\sim~{<M>~\nib\over R^{3}(t_{in})},\eqno(4.18)
$$
where $<M>\sim 10^{14}~g$ and
$$
R(t_{in})~\sim~R_{0}~(t_{0}/t_{in})^{\alpha};\eqno(4.19)
$$
here $R_{0}=1.25\times 10^{28}~cm~=1.4\times 10^{10}~lyr$ in a typical
cosmological model [27] while for $\alpha$ we can take 
the value $-0.5$. 
Then, if one assumes for the PBHs density a behaviour that approximately 
scales as a power with exponent $2/3$ of the time and 
an initial time $t_{in}\sim 10^{-70}\div 10^{-75}~ sec$,
the density parameter $\Omega_{PBH}$
at the present time is:
$$
\Omega_{PBH}~=~ {\rho_{0PBH} \over \rho_{cr}}~=~
1.12\times 10^{-12}~\div 1.65\times 10^{-8};
\eqno(4.20)
$$
In tab. 4 I listed the results obtained by choosing 
different formation times for the primordial black holes:
this rough estimate is purely indicative but 
in any case it may reproduce quite well the present experimental upper limit
of the density parameter,
derived from the constraints imposed by studying the CMB characteristics 
[10]:
$
\Omega_{PBH}\leq
(7.6\pm 2.6)\times 10^{-9}~h^{(-1.95\pm 0.15)}.
$
\vskip 7mm
\centerline{{\bf 5. RESULTS AND CONCLUSIONS.}}
\vskip 7mm
In figs. 1-6 I plotted the behaviour of the ionization degree $x$ 
vs $-z$, as
results from the solution of eq. (4.16a): the six different plots 
refer to values of the reionization redshift $\zre$ ranging from 15 to 60;
in figs. 7-12 I plotted
the evolution vs $-z$ of the plasma temperature;
for convenience, I also reported 
in figs. 13-18 and 19-24 the corresponding results obtained
in Ref. [6] by using an exact blackbody spectrum without jets emission.

The present results should be compared with the ones obtained in [6]
by a first order solution of the differential equation system;
we can remark that:

\item {a)} as in Ref. [6], the black holes-induced reionization
is only partial; a quite relevant effect is obtained for 
an evaporation redshift corresponding to 
$\zre \leq 30$, while for higher values of $\zre$ the process
of PBHs quantum evaporation cannot produce an appreciable 
phenomenon of reionization; in fact, for $\zre \geq 40$ the
ionization degree is $x\leq 0.35$.
However, with respect to the results of Ref. [6],
a smaller density of PBHs is necessary to reionize the Universe because 
in this case, one adds to the 
direct photon emission the indirect one, due to the hadrons decays.
Moreover, the plasma heating is limited by a powerful
cooling: the overall effect is that in eq. (3.1) the $\ppi$ term is 
suppressed and the $\tri$ term enhanced.

\item{b)} Discussing now the result obtained for the plasma temperature and
looking at figs. 7-12, one can remark that a real improvement of the
calculation has been obtained: the consideration of all
the cooling terms without any approximations enables me to have a maximum 
plasma temperature of the order of the tenth of eV
while previously the PBHs quantum evaporation was producing 
an excessive plasma heating; the irregularities in the plots are 
due to some numerical fluctuations attributable to 
the routine solving the coupled differential equations system.

These fluctuations are more evident when a shorter integration time 
is considered. 

\item {c)} Looking at both the results for the ionization degree and
the plasma temperature, one can conclude that this PBHs-induced 
reionization mechanism should effectively be ``late and sudden'':
the possibility of an effective reionization 
for values $\zre > 30$ should be discarded and thus, in this model,
it is very improbable to have a significative suppression of
the CMB temperature fluctuations on small angular scales as predicted by 
Bond and Efstathiou [28] and
Vittorio and Silk [29]; in these works, one proves 
that an early reionization could suppress in a significant 
way these fluctuations, that, following the predictions 
of CDM models and texture scenarios [30], result too large.

\item {d)} The PBHs formation time should be put very far in the past:
considering the jet emission contribution in the photon spectrum,
the best choice of 
the PBHs formation conditions that enables to obtain a well 
balanced reionization process (i.e. an high ionization degree without 
an excessive plasma heating) suggests a 
formation time in a range $t_{form}=10^{-70}~\div~10^{-75}~sec$ after
the Big Bang and an initial density $\rho_{i}=10^{17}~\div~10^{24}g/cm^{3}$
(see tab. 4), corresponding to a 
present density parameter $\Omega_{PBH}\sim 1.12\times 10^{-12}$ 
$\div~1.65\times 10^{-8}$.
\vskip 7mm
Summarizing, in this paper I studied a mechanism of reionization for 
the Universe induced by the quantum evaporation of primordial
black holes.
Differently from a previous analysis, here an emission spectrum taking 
into account the quark and gluons jets production has been considered.

A most accurate solution of
the system of coupled differential equations, giving the ionization degree and 
the plasma temperature evolution with the time,
is performed; as a result, the reionization of the Universe happens
in an effective, even if partial, way. In the same time, the 
resulting plasma heating is limited by 
an effective cooling, due to excitation and 
collisional processes; that avoids to have large
(experimentally unseen) distortions of the CBR blackbody spectrum
via the Sunyaev- Zel'dovich effect.

As in Ref. [6], I considered here some values of the reionization
redshift in the range [15, 60]: the fast rise of the ionization degree $x$
seems to suggest that
such a model of PBHs-induced reionization
should be classified as ''sudden and late''. 
In fact, the behaviour of $x$ can be approximated by an exponential
function: this result justifies the analysis of Ref. [31],
where I studied the consequences of an exponential reionization 
of the Universe
on the polarization of the Cosmic Microwave Background. 

Finally, the possibility to have a total reionization for a 
redshift $\zre > 60$ seems to be excluded by the results here found: 
in this model, no damping of the temperature fluctuations
at small angular scales should be expected.
\vfill\eject
\centerline{{\bf ACKNOWLEDGEMENTS}}
\vskip 7mm
I would like to thank the Universit\'a degli Studi di Milano for its financial
support and the Queen Mary and Westfield College for its hospitality
and for the technical support given to my work. 
I am grateful to Bernard Carr and Peter Coles for many
useful discussions about the reionization problems and the black holes
physics; in particular, the works of B. Carr 
have been fundamental for my research.

I am really grateful to all the people of the Astrophysics Section 
of the University of Milano, but a particular thank goes to my Tutor,
Silvio Bonometto, for his fundamental suggestions and for the continuous 
scientific support he gave me.

Finally, I want to gratefully thank Ruth Durrer,
for the useful references she sent me.
\vskip 1cm
\centerline {\bf{REFERENCES}}
\vskip 7mm

\item{[1]} Gunn, J.E., Peterson, B.A., Astroph. J.,{\bf 142}, 1633, (1965).

\item{[2]} Stebbins, A., Silk, J., Astroph. J.,{\bf 300}, 1, (1986);

\item{[3]} Arons, J., Wingert, D., Astroph. J.,{\bf 177}, 1, (1972).

\item{} Ginzburg, V., Ozernoi, L., Soviet Astr., {\bf 9}, 726, (1966).

\item{[4]} Tuluie, R., Matzner, R.A., Anninos, P., ''Anisotropies of
the Cosmic Background Radiation in a Reionized Universe'', Preprint,
Unpublished.

\item{[5]} Fukugita, M., Kawasaki, M., Astroph. J.,{\bf 353}, 384, (1990).

\item{} Gabbiani, F.,Masiero, A., Sciama, D.W., Phys. Lett., {\bf B259}, 323,
(1991).

\item{[6]} Gibilisco, M: ''Reionization of the Universe induced by 
Primordial Black Holes'', submitted to Int. Journ. of Mod. Phys A, Jan. 1996.

\item{[7]} Hawking, S.W., Commun. Math. Phys., {\bf 43}, 199, (1975).

\item{[8]} Mather, J.C. et al., Astroph. J., {\bf 420}, 439, (1994);

\item{[9]} Sunyaev, R.A., Zel'dovich, Y.B., Commun. Astroph. Sp. Phys., 
{\bf 4}, (1973), 173; and Ann. Rev. Astron. Astroph., {\bf 18}, (1980), 537.

\item{[10]} Mac Gibbon, J.H., Carr, B.J., Astroph. J., {\bf 371}, 447, (1991);

\item {} Mac Gibbon, J.H., Webber, B.R., Phys. Rev., {\bf D41}, 3052, (1990).
                                         
\item{[11]} Damour, T. and Ruffini, R., Phys. Rev. Lett., {\bf 35}, 463, 
(1975);

\item {} Ternov, I.M., Gaina, A.B., Chiznov, G. A., Sov. J. Nucl. 
Phys., {\bf 44}, 343, (1986);

\item {}Page, D., Phys. Rev., {\bf D14}, 3260, (1976).
                             
\item{[12]} Hiscock, W.A., Weems, L. D., Phys. Rev., {\bf D41}, 1142, (1990).

\item{[13]} Page, D. N., Phys. Rev., {\bf D16}, 2402, (1977).
                                               
\item{[14]} Novikov, I., ''Black Holes and the Universe'', Cambridge Univ.
Press, 1990.

\item {[15]} Hawking, S.W., Mon. Not. R. Astron. Soc., {\bf 152}, 75, (1971).

\item{[16]} Carr, B.J., Lidsey, J.E., Phys. Rev., {\bf D48}, 543, (1993).

\item{[17]} Hawking, S.W. et al., Phys. Rev., {\bf D26}, 2681, (1982);

\item{} La, D., Steinhardt, P.J., Phys. Lett, {\bf B220}, 375, (1989).

\item{[18]} Khlopov, M.Y., Polnarev, A.G., Phys. Lett., {\bf B97}, 383, (1980).

\item{[19]} Hawking, S.W., Phys. Lett., {\bf B231}, 237, (1989);

\item{} Polnarev, A.G., Zemboricz, R., Phys. Rev., {\bf D43}, 1106, (1991).

\item{[20]} Mac Gibbon, J. H., Phys. Rev., {\bf D44}, 376, (1991).

\item{[21]} Page, D. N., Phys. Rev., {\bf D13}, 198, (1976).

\item{[22]} Carr, B. J., Astronomical and Astroph. Transactions, Vol. {\bf 5},
43, (1994).

\item{[23]} Durrer, R., Infrared Phys. Technol., {\bf 35}, (1994), 83.

\item{[24]} Peebles, P.J.E., Astroph. J.,{\bf 153}, 1, (1968).

\item{[25]} Peebles, P.J.E., ''Principles of Physical Cosmology'', p. 166-177,
Princeton University Press, 1993.

\item{[26]} Sneddon, I.A.:'' Elements of partial differential equations'',
McGraw-Hill, N.Y., 1957.

\item{[27]} Misner, C. W., Thorne, K.S., Wheeler, J. A.: ''Gravitation'',
p. 738, W. H. Freeman and Co., San Francisco, (1973).

\item{[28]} Bond, J.R., Efstathiou, G., Astroph. J.,{\bf 285}, L45, (1984).

\item{[29]} Vittorio, N., Silk, J., Astroph. J.,{\bf 285}, L39, (1984).

\item {[30]} Turok, N., Phys. Rev. Lett., {\bf 63}, 2625, (1989);
                        
\item{} Turok, N., Spergel, D.N., Phys. Rev. Lett., {\bf 64}, 2736 (1990);

\item{} Durrer, R., Phys. Rev., {\bf D42}, 2533, (1990).

\item {} Tegmark, M., Silk, J., ''On the inevitability of Reionization:
Implications for Cosmic Microwave Background Fluctuations'', Preprint
CfPA-93-th-04, June 1993.

\item{[31]} Gibilisco, M.,
Intern. Journal of Modern Phys, {\bf 10A}, 3605, (1995).

\vfill\eject
                                                        
{\bf Tab. 1: Maximum energy for photons affected by ionization losses 
before a redshift $\tilde z$.}
\vskip 0.5cm
{\offinterlineskip
\tabskip=0pt
\halign{ \strut
	 \vrule#&
\quad	 \hfil # &
	 \vrule# &
\quad	 \hfil # &
	 \vrule#
	 \cr
\noalign{\hrule}
&                 &&                                         &\cr
& $\tilde z$      && $\omega_{max}~(\times 10^{-6})~GeV$~~~~ &\cr
&                 &&                                         &\cr
\noalign{\hrule}
&                 &&                                         &\cr
& 15              && ~~~~~~~~~~~5.54~~~~~~~~~~~~~~~          &\cr
&                 &&                                         &\cr
& 20              && ~~~~~~~~~~~6.00~~~~~~~~~~~~~~~          &\cr
&                 &&                                         &\cr
& 30              && ~~~~~~~~~~~6.70~~~~~~~~~~~~~~~          &\cr
&                 &&                                         &\cr
& 40              && ~~~~~~~~~~~7.20~~~~~~~~~~~~~~~          &\cr
&                 &&                                         &\cr
& 50              && ~~~~~~~~~~~7.70~~~~~~~~~~~~~~~          &\cr
&                 &&                                         &\cr
& 60              && ~~~~~~~~~~~8.10~~~~~~~~~~~~~~~          &\cr
&                 &&                                         &\cr
\noalign{\hrule}
}}
                                                                             
\vfill\eject

{\bf Tab. 2: Time averaged values of the integrated photon density
$A_{1},~B_{1}$.}

\vskip 1cm 
\vskip 7mm
{\offinterlineskip
\tabskip=0pt
\halign{ \strut
	 \vrule#&
\quad	 \hfil # &
	 \vrule# &
\quad	 \hfil # &
	 \vrule# &
\quad	 \hfil # &
	 \vrule# &
\quad	 \hfil # &
	 \vrule#
	 \cr
\noalign{\hrule}
&                 &&            &&             &&                    &\cr
& $z_{R}$         && $\try$~~~  && $\as1$~~~   && $\bs1$ ~~~         &\cr
&                 &&            &&             &&                    &\cr
\noalign{\hrule}
&                 &&            &&             &&                    &\cr
& 15              && 6.43 $\et$ && 10.4 $\ea$  && 3.83 $\eb$         &\cr
&                 &&            &&             &&                    &\cr
& 20              && 4.30 $\et$ && 8.88 $\ea$  && 3.54 $\eb$         &\cr
&                 &&            &&             &&                    &\cr
& 30              && 2.38 $\et$ && 7.12 $\ea$  && 3.17 $\eb$         &\cr
&                 &&            &&             &&                    &\cr
& 40              && 1.57 $\et$ && 6.17 $\ea$  && 2.96 $\eb$         &\cr
&                 &&            &&             &&                    &\cr
& 50              && 1.13 $\et$ && 5.40 $\ea$  && 2.76 $\eb$         &\cr
&                 &&            &&             &&                    &\cr
& 60              && 0.86 $\et$ && 4.88 $\ea$  && 2.63 $\eb$         &\cr
&                 &&            &&             &&                    &\cr
\noalign{\hrule}
}}
                                                                             
\vfill\eject                                                          
                                                                     
{\bf Tab. 3: Value of the coefficients $C_{1}(z_{R})$, $C_{2}(z_{R})$
respectively appearing in $t_{PI}^{-1}$ and $\Gamma_{PI}$.}
\vskip 1cm
{\offinterlineskip
\tabskip=0pt
\halign{ \strut
	 \vrule#&
\quad	 \hfil # &
	 \vrule# &
\quad	 \hfil # &
	 \vrule# &
\quad	 \hfil # &
	 \vrule#
	 \cr
\noalign{\hrule}
&                 &&                  &&                                 &\cr
& $z_{R}$         &&  $C_{1}$~~~~     &&  $C_{2}$~~~~                    &\cr
&                 &&                  &&                                 &\cr
\noalign{\hrule}
&                 &&                  &&                                 &\cr
& 15              && 8.20 $\qa$       &&  3.02 $\qb$                     &\cr
&                 &&                  &&                                 &\cr
& 20              && 6.99 $\qa$       &&  2.79 $\qb$                     &\cr
&                 &&                  &&                                 &\cr
& 30              && 5.61 $\qa$       &&  2.50 $\qb$                     &\cr
&                 &&                  &&                                 &\cr
& 40              && 4.86 $\qa$       &&  2.33 $\qb$                     &\cr
&                 &&                  &&                                 &\cr
& 50              && 4.25 $\qa$       &&  2.18 $\qb$                     &\cr
&                 &&                  &&                                 &\cr
& 60              && 3.84 $\qa$       &&  2.07 $\qb$                     &\cr
&                 &&                  &&                                 &\cr
\noalign{\hrule}
}}
\vfill\eject                          
                                                                         
{\bf Tab. 4: Values of the density parameter $\Omega_{PBH}$ and of 
the PBHs number densities for different choices of the initial 
black holes birth time $t_{i}$ (Exp: $\Omega_{PBH}\leq(7.6\pm 2.6)\times
10^{-9}h^{(-1.95\pm 0.15)}$).}
\vskip 7mm
{\offinterlineskip
\tabskip=0pt
\halign{ \strut
	 \vrule#&
\quad	 \hfil # &
	 \vrule# &
\quad	 \hfil # &
	 \vrule# &
\quad	 \hfil # &
	 \vrule# &
\quad	 \hfil # &
	 \vrule#
	 \cr
\noalign{\hrule}
&                 &&            &&             &&                    &\cr
& $t_{i}~(sec)$   && $\rho_{i}~(g~cm^{-3})$      
                  && $\rho_{0}~(g~cm^{-3})$         
                  && $\Omega_{PBH}$                                  &\cr
&                 &&            &&             &&                    &\cr
\noalign{\hrule}
&                 &&            &&             &&                    &\cr
& $10^{-75}$      && $4.28\times 10^{24}~$
		  && $7.74\times 10^{-38}~$   
                  && $1.65\times 10^{-8}~$                           &\cr
&                 &&            &&             &&                    &\cr
& $10^{-73}$      && $4.28\times 10^{21}~$
		  && $1.66\times 10^{-39}~$   
                  && $3.53\times 10^{-10}~$                           &\cr
&                 &&            &&             &&                    &\cr
& $10^{-71}$      && $4.28\times 10^{18}~$
		  && $3.59\times 10^{-41}~$   
                  && $7.64\times 10^{-12}~$                          &\cr
&                 &&            &&             &&                    &\cr
& $10^{-70}$      && $1.35\times 10^{17}~$
		  && $5.26\times 10^{-42}~$   
                  && $1.12\times 10^{-12}~$                          &\cr
&                 &&            &&             &&                    &\cr
\noalign{\hrule}
}}
                                        
\vfill\eject
{\bf FIGURE CAPTIONS}
\vskip 1cm
\item {Fig. 1:} Evolution of the ionization degree $x$ vs $-z$ induced by 
primordial black holes evaporation; the Hawking emission spectrum 
is corrected for the presence of quarks and gluons jets; the plot
is obtained by assuming a reionization redshift $z=15$.

\vskip 3mm
\item {Figs. 2-6:} The same evolution of $x$ but, respectively, 
for $z=20$, $z=30$, $z=40$, $z=50$, $z=60$.

\vskip 3mm
\item {Fig. 7:} The evolution of the plasma temperature $T_{e}$ 
vs $-z$; the reionization redshift is $z=15$.

\vskip 3mm
\item {Fig. 8-12:} The same evolution of $T_{e}$ but, respectively, 
for $z=20$, $z=30$, $z=40$, $z=50$, $z=60$.

\vskip 3mm
\item {Fig. 13:} The evolution of the ionization degree $x$ is calculated
by using an exact blackbody emission spectrum in a first order recursive
calculation (see Ref. 6) and plotted as a function of $-z$;
the reionization redshift considered is $z=15$.

\vskip 3mm
\item {Figs. 14-18:} The same evolution of $x$ but, respectively, 
for $z=20$, $z=30$, $z=40$, $z=50$, $z=60$.

\vskip 3mm
\item {Fig. 19:} The evolution of the plasma temperature $T_{e}$ is calculated
by using an exact blackbody emission spectrum in a first order recursive
calculation (see Ref. 6) and plotted as a function of $-z$;
the reionization redshift considered is $z=15$.

\vskip 3mm
\item {Figs. 20-24:} The same evolution of $x$ but, respectively, 
for $z=20$, $z=30$, $z=40$, $z=50$, $z=60$.

\end